\documentclass[apj]{emulateapj_arxiv}
\usepackage{graphicx}
\usepackage{color}
\usepackage{amsmath}

\begin{document}
\title{A Proposed Paradigm for Solar Cycle Dynamics Mediated via Turbulent Pumping of Magnetic Flux in Babcock-Leighton type Solar Dynamos}

\author{Soumitra Hazra\altaffilmark{1,2} and Dibyendu Nandy \altaffilmark{1,2}}
\affil{$^1$Department of Physical Sciences, Indian Institute of Science Education and Research, Kolkata}
\affil{$^2$Center of Excellence in Space Sciences India, IISER Kolkata, Mohanpur 741246, West Bengal, India}
\begin{abstract}

At present, Babcock-Leighton flux transport solar dynamo models appear as the most promising model for explaining diverse observational aspects of the sunspot cycle. The success of these flux transport dynamo models is largely dependent upon a single-cell meridional circulation with a deep equatorward component at the base of the Sun's convection zone. However, recent observations suggest that the meridional flow may in fact be very shallow (confined to the top 10\% of the Sun) and more complex than previously thought. Taken together these observations raise serious concerns on the validity of the flux transport paradigm. By accounting for the turbulent pumping of magnetic flux as evidenced in magnetohydrodynamic simulations of solar convection, we demonstrate that flux transport dynamo models can generate solar-like magnetic cycles even if the meridional flow is shallow. Solar-like periodic reversals is recovered even when meridional circulation is altogether absent, however, in this case the solar surface magnetic field dynamics does not extend all the way to the polar regions. Very importantly, our results demonstrate that the Parker-Yoshimura sign rule for dynamo wave propagation can be circumvented in Babcock-Leighton dynamo models by the latitudinal component of turbulent pumping -- which can generate equatorward propagating sunspot belts in the absence of a deep, equatorward meridional flow. We also show that variations in turbulent pumping coefficients can modulate the solar cycle amplitude and periodicity. Our results suggest the viability of an alternate magnetic flux transport paradigm -- mediated via turbulent pumping -- for sustaining solar-stellar dynamo action.

\end{abstract}

\section{Introduction}
The cycle of sunspots involves the generation and recycling of the Sun's toroidal and poloidal magnetic field components. The magnetohydrodynamic (MHD) dynamo mechanism that achieves this is sustained by the energy of solar internal plasma motions such as differential rotation, turbulent convection and meridional circulation. The toroidal field is generated through stretching of the poloidal component by differential rotation \cite{park55} and is believed to be stored and amplified at the overshoot layer \cite{moren92} beneath the base of the solar convection zone (SCZ). Strong toroidal flux tubes are unstable to magnetic buoyancy and erupt through the surface producing sunspots, which are strongly magnetized and have a systematic tilt \cite{hale08,hale19}. The poloidal field is believed to be regenerated through a combination of helical turbulent convection (traditionally known as the mean-field $\alpha$-effect; \cite{park55}) in the main body of the SCZ and the redistribution of the magnetic flux of tilted bipolar sunspot pairs (the Babcock-Leighton process; \cite{bab61,leigh69}).

Despite early, pioneering attempts to self-consistently model the interactions of turbulent plasma flows and magnetic fields in the context of the solar cycle \cite{gilm83, glat85} such full MHD simulations are still not successful in yielding solutions that can match solar cycle observations. This task is indeed difficult, for the range of density and pressure scale heights, scale of turbulence and high Reynolds number that characterize the SCZ is difficult to capture even in the most powerful supercomputers. An alternative approach to modelling the solar cycle is based on solving the magnetic induction equation in the SCZ with observed plasma flows as inputs and with additional physics gleaned from simulations of convection and flux tube dynamics. These so called flux transport dynamo models have shown great promise in recent years in addressing a wide variety of solar cycle problems \cite{char10, ossen03}.

In particular, solar dynamo models based on the Babcock-Leighton mechanism for poloidal field generation have been more successful in explaining diverse observational features of the solar cycle \cite{dikp99, nandy02, chat04, chou04, guer07, nandy11, chou12, haz14, pass14}. Recent observations also strongly favor the Babcock-Leighton mechanism as a major source for poloidal field generation \cite{dasi10, munoz13}. In this scenario, the poloidal field generation is essentially predominantly confined to near-surface layers. For the dynamo to function efficiently, the toroidal field that presumably resides deep in the interior has to reach the near-surface layers for the Babcock-Leighton poloidal source to be effective. This is achieved by the buoyant transport of magnetic flux from the Sun's interior to its surface (through sunspot eruptions). Subsequent to this the poloidal field so generated at near-surface layers must be transported back to the solar interior, where differential rotation can generate the toroidal field. The deep meridional flow assumed in such models (See Fig.~1, left-hemisphere) plays a significant role in this flux transport process and is thought to govern the period of the sunspot cycle \cite{char20, hatha03, yeat08, ghaz14}. Moreover, a fundamentally crucial role attributed to the deep equatorward meridional flow is that it allows the Parker-Yoshimura sign rule \cite{park55, yosh75} to be overcome, which would otherwise result in poleward propagating dynamo waves in contradiction to observations that the sunspot belt migrates equatorwards with the progress of the cycle \cite{chou95,ghaz14,pass15, belus15}.

While the poleward meridional flow at the solar surface is well observed (Hathaway \& Rightmire 2010; 2011) the internal meridional flow profile has remained largely unconstrained. A recent study utilizing solar supergranules \cite{hatha12} suggests that the meridional flow is confined to within the top 10\% of the Sun (Fig.~1, right-hemisphere) -- much shallower than previously thought. Independent studies utilizing helioseismic inversions are also indicative that the equatorward meridional counterflow may be located at shallow depths \cite{mitra07, zhao13}. The latter also infer the flow to be multi-cellular and more complex. These studies motivate exploring alternative paradigms for flux transport dynamics in Babcock-Leighton type models of the solar cycle which are crucially dependent on meridional circulation linking the two segregated dynamo source regions in the SCZ. This leads us to consider the role of turbulent pumping.
\begin{figure}[!htb]
  \begin{center}
\begin{tabular}{cc}
\includegraphics[scale=0.43]{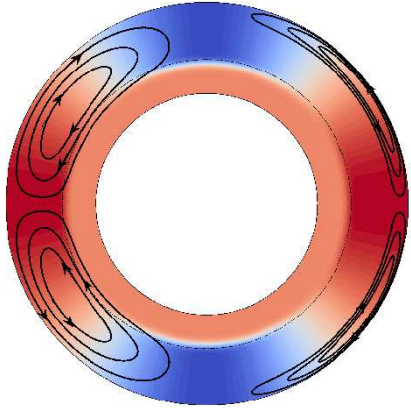}
\end{tabular}
\end{center}
\caption{The outer 45\% of the Sun depicting the internal rotation profile in color. Faster rotation is denoted in deep red and slower rotation in blue. The equator of the Sun rotates faster than the polar regions and there is a strong shear layer in the rotation near the base of the convection zone (denoted by the dotted line). Streamlines of a deep meridional flow (solid black curves) reaching below the base of the solar convection zone (dashed line) is shown on the left hemisphere, while streamlines of a shallow meridional flow confined to the top 10\% of the Sun is shown on the right hemispheres (arrows indicate direction of flow). Recent observations indicate that the meridional flow is much shallower and more complex than traditionally assumed, calling in to question a fundamental premise of flux transport dynamo models of the solar cycle.}
\end{figure}
Magnetoconvection simulations supported by theoretical considerations  have established that turbulent pumping preferentially transports magnetic fields vertically downwards \cite{brand96, tobias01, ossen02, dorch01, kapla06, pip09, racine11, rogach11, aug15, warn16, sim16} -- likely mediated via strong downward convective plumes which are particularly effective on weak magnetic fields (such as the poloidal component). In strong rotation regimes, there is also a significant latitudinal component of turbulent pumping. In particular, two studies, one utilizing mean-field dynamo simulations \cite{brand92} and the other utilizing turbulent three dimensional magnetoconvection simulations \cite{ossen02} recognized the possibility that turbulent pumping may contribute to the equatorward propagation of the toroidal field belt. We note that most Babcock-Leighton kinematic flux transport solar dynamo models do not include the process of turbulent pumping of magnetic flux. The few studies that exist on the impact of turbulent pumping in the context of flux transport dynamo models show it to be dynamically important in flux transport dynamics, the maintenance of solar-like parity and solar-cycle memory \cite{guer08, kar12, jiang13}. In their model with turbulent pumping, Guerrero \& de Gouveia Dal Pino (2008) used a spatially distributed $\alpha$-coefficient in the near-surface layers to model the Babcock-Leighton poloidal source and a meridional circulation whose equatorward component penetrated up to $0.8R_\odot$, i.e., more than half the depth of the SCZ; therefore, from this modelling it is not possible to segregate the contributions of turbulent pumping and meridional flow (the peak latitudinal component of the former coincides with the equatorward component of the latter) to the toroidal field migration.

Here, utilizing a newly developed state-of-the-art flux transport dynamo model where a double-ring algorithm is utilized to model the Babcock-Leighton process, we explore the impact of turbulent pumping in flux transport dynamo models with nonexistent, or shallow meridional circulation. Our results indicate the possibility of an alternative flux transport paradigm for the solar cycle in which turbulent pumping of magnetic flux resolves the problems posed by a shallow (or inconsequential) meridional flow.
\begin{figure}[!htb]
  \begin{center}
\begin{tabular}{cc}
\includegraphics[scale=0.7]{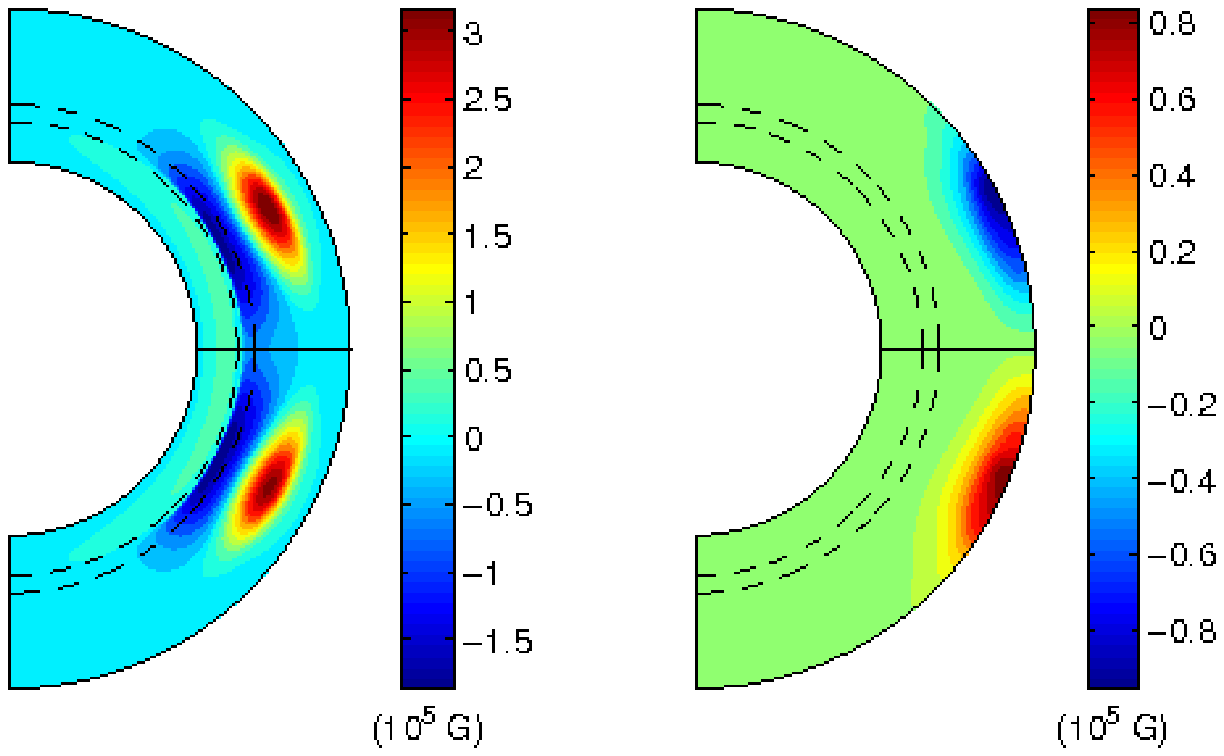} \\
(a)~~~~~~~~~~~~~~~~~~~~~~~~~~~~~~~~~~~~~~~~~~(b)\\
\includegraphics[scale=0.7]{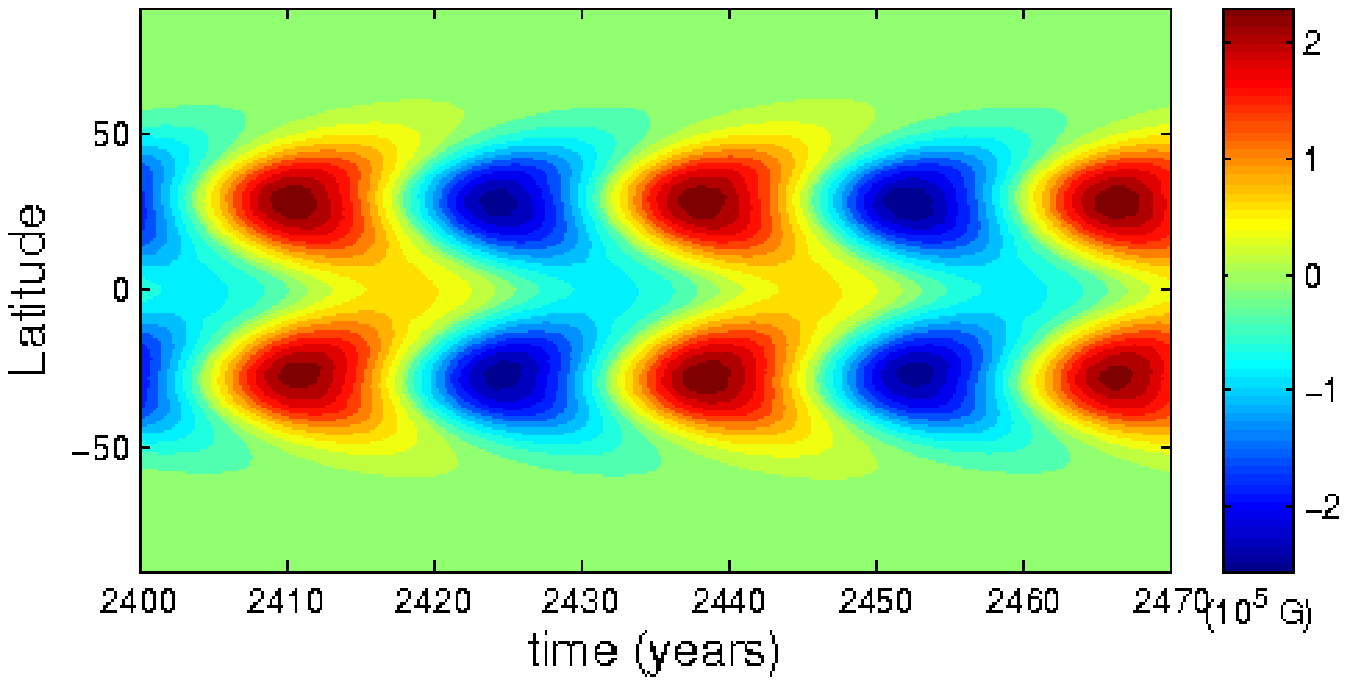} \\
~~~~~~(c)
 \end{tabular}
\end{center}
\caption{Solar cycle simulations with a shallow meridional flow. The toroidal (a) and poloidal (b) components of the magnetic field is depicted within the computational domain at a phase corresponding to cycle maxima. The solar interior shows the existence of two toroidal field belts, one at the base of the convection zone and the other at near-surface layers where the shallow equatorward meridional counterflow is located. Region between two dashed circular arcs indicates the tachocline. (c) A butterfly diagram generated at the base of convection zone showing the spatiotemporal evolution of the toroidal field. Latitude are in degrees. Clearly, there is no dominant equatorward propagation of the toroidal field belt and the solution displays quadrupolar parity (i.e., symmetric toroidal field across the equator) which do not agree with observations.}
\end{figure}
\section{Model}
Our flux transport solar dynamo model solves for the coupled, evolution equation for the axisymmetric toroidal and poloidal components of the solar magnetic fields:
\begin{equation}\label{Eq_2.5DynA}
    \frac{\partial A}{\partial t} + \frac{1}{s}\left[ \textbf{v}_p \cdot \nabla (sA) \right] = \eta\left( \nabla^2 - \frac{1}{s^2}  \right)A + {S_{BL}},
\end{equation}

\begin{eqnarray}\label{Eq_2.5DynB}
    \frac{\partial B}{\partial t}  + s\left[ \textbf{v}_p \cdot \nabla\left(\frac{B}{s} \right) \right]
    + (\nabla \cdot \textbf{v}_p)B = \eta\left( \nabla^2 - \frac{1}{s^2}  \right)B  \nonumber \\
    + s\left(\left[ \nabla \times (A\bf \hat{e}_\phi) \right]\cdot \nabla \Omega\right)
    + \frac{1}{s}\frac{\partial (sB)}{\partial r}\frac{\partial \eta}{\partial r},~~~~~
\end{eqnarray}
where, $B$  is the toroidal component of magnetic field and $A$ is the vector potential for the poloidal component of magnetic field. ${\textbf v}_p$ is the meridional flow, $\Omega$ is the differential rotation, $\eta$ is the turbulent magnetic diffusivity and $s = r\sin(\theta)$. For the differential rotation and diffusivity profile, we use an analytic fit to the observed solar differential rotation (the near-surface shear layer is not included) and a two-step turbulent diffusivity profile (which ensures a smooth transition to low levels of diffusivity beneath the base of the convection zone) (For detailed profile, see Hazra \& Nandy 2013). We use the same meridional flow profile as defined in Hazra \& Nandy (2013). Our flow profile has penetration depth of $0.65R_\odot$ to represent deep meridional flow situation, and $0.90~R_\odot$ to represent shallow meridional flow situation. We set the peak speed of the meridional flow to be 15 ms$^{-1}$ (near mid-latitudes). The second term on the RHS of the toroidal field evolution equation acts as the source term for the toroidal field (rotational shear), while in the poloidal field evolution equation, the source term, ${S_{BL}}$, is due to the Babcock-Leighton mechanism. Here we use a double-ring algorithm for buoyant sunspot eruptions that best captures the Babcock-Leighton mechanism for poloidal field generation \cite{durney97, nandy01, munoz10,haz13} and which has been tested thoroughly in other contexts. Specifics about our double ring algorithm can be found in Hazra \& Nandy (2013) and Hazra (2016; PhD Thesis).
\begin{figure}[!htb]
\begin{center}
\begin{tabular}{cc}
\includegraphics[scale=0.7]{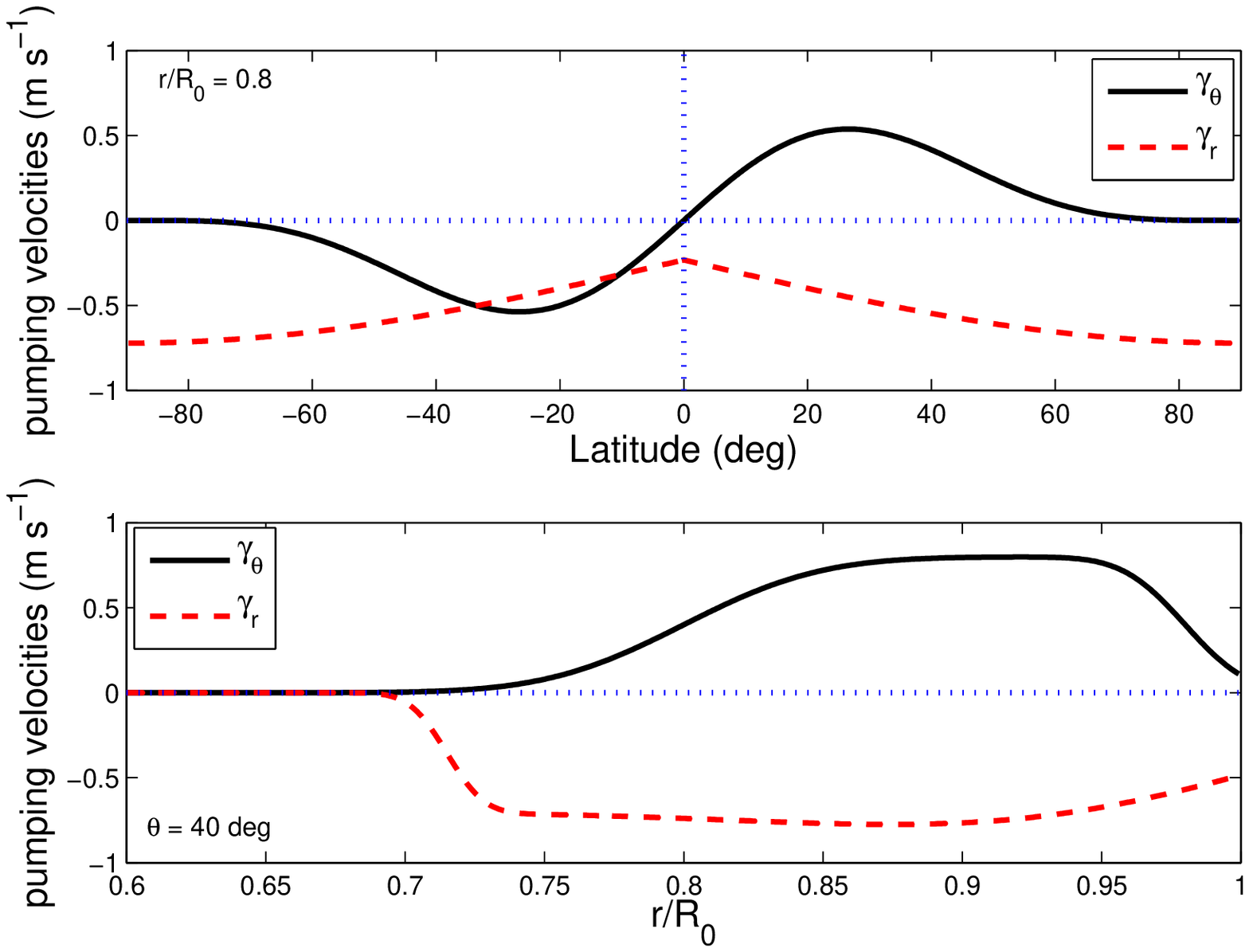}

 \end{tabular}
\end{center}
\caption{Latitudinal (top) and radial (bottom) variation of the radial (dashed lines) and latitudinal (solid lines) turbulent pumping velocity components taken at a depth of 0.8 $R_{\odot}$ (top plot) and at a colatitude $40^\circ$ (bottom plot). Radial turbulent pumping is negative (downward) in both hemispheres. Latitudinal turbulent pumping is equatorward throughout the convection zone in both the hemispheres.}
 \end{figure}
\section{Results}

To bring out the significance of the recent observations, we first consider a single cell, shallow meridional flow, confined only to the top 10\% of the convection zone (Fig.~1, right-hemisphere). In the first scenario we seek to answer the following question: Can solar-like cycles be sustained through magnetic field dynamics completely confined to the top 10\% of the Sun?

In these simulations initialized with antisymmetric toroidal field condition (with initial B $\sim$ 100 kG), we first allow magnetic flux tubes to buoyantly erupt from 0.90 $R_{\odot}$ (i.e., the depth to which the shallow flow is confined) when they exceed a buoyancy threshold of $10^4$ Gauss (G). In this case we find that the simulated fields  fall and remain below this threshold (at all latitudes at 0.90 $R_{\odot}$) with no buoyant eruptions, implying that a Babcock-Leighton type solar dynamo cannot operate in this case. Dikpati {\it et al.} (2002) considered the contribution of the near-surface shear layer in their simulations (which we have not) and concluded that this near-surface layer contributes only about 1 kG to the total toroidal field production and hence insufficient to drive a large-scale dynamo. Guerrero \& de Gouveia Dal Pino (2008) also utilized a near-surface shear layer with radial pumping and found solar-like solutions only under special circumstances; however, given that for this particular case they utilized a local $\alpha$-effect for the latter simulations (with a spatially distributed $\alpha$-effect in the near-surface layer) it is not evident that these simulations are relatable to the Babcock-Leighton solar dynamo concept. The upper layers of the SCZ is highly turbulent and storage and amplification of strong magnetic flux tubes may not be possible in these layers \cite{park75, moren83} and therefore this result is not unexpected. While Brandenburg (2005) has conjectured that the near-surface shear layer may be able to power a large-scale dynamo, this remains to be convincingly demonstrated in the context of a Babcock-Leighton dynamo.

In the second scenario with a shallow meridional flow, we allow magnetic flux tubes to buoyantly erupt from 0.71 $R_{\odot}$, i.e. from base of the convection zone. In this case we get periodic solutions but analysis of the butterfly diagrams (taken both at the base of SCZ and near solar surface) shows that the toroidal field belts have almost symmetrical poleward and equatorward branches with no significant equatorward migration (see Fig.~2). Moreover, as already noted by Guerrero, G. \& de Gouveia Dal Pino (2008), the solutions with shallow meridional flow always display quadrupolar parity in contradiction with solar cycle observations. Clearly, a shallow flow poses a serious problem for solar cycle models.
\begin{figure}[!htb]
\begin{center}
\begin{tabular}{cc}
\includegraphics[scale=0.7]{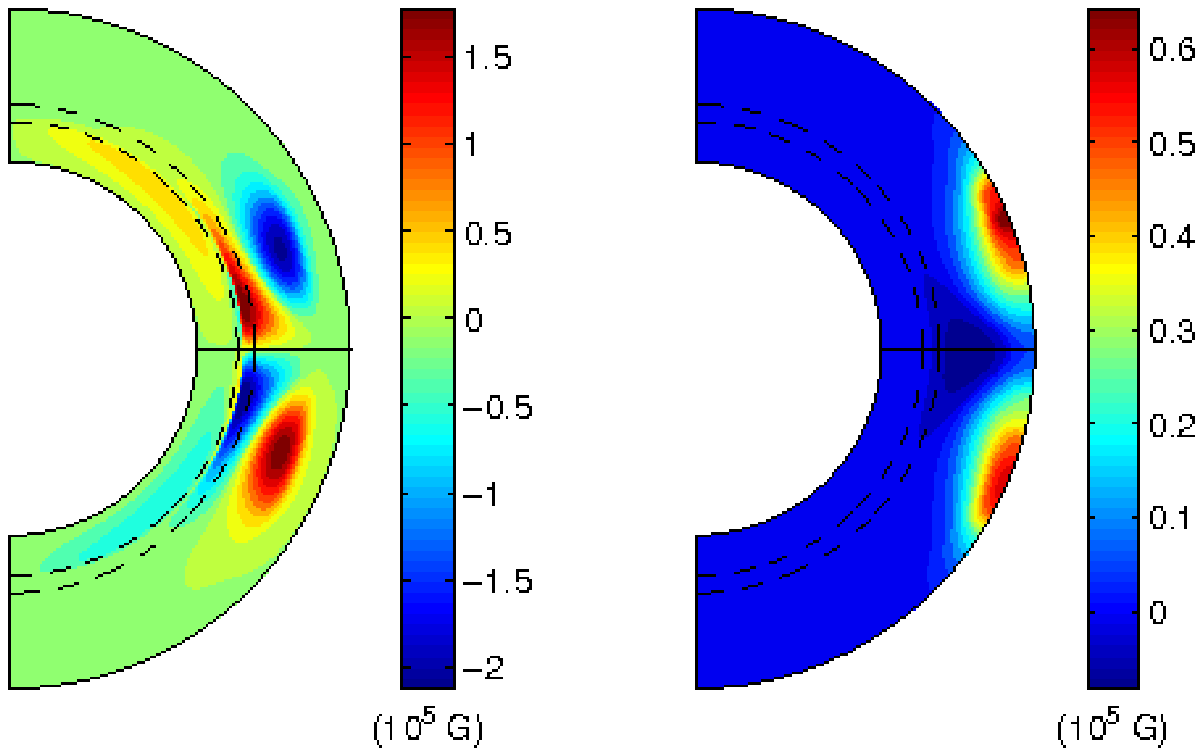} \\
(a)~~~~~~~~~~~~~~~~~~~~~~~~~~~~~~~~~~~~~~~~~~~~~~~(b)\\
\includegraphics[scale=0.7]{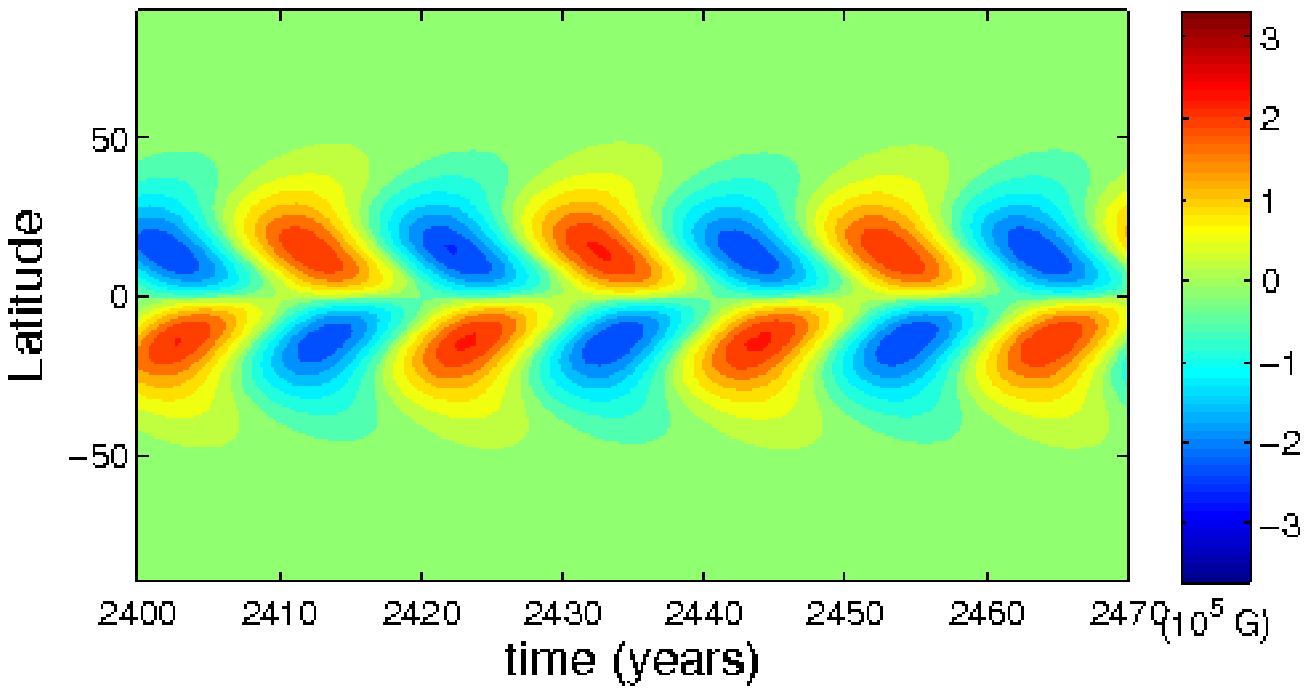} \\
~~~~(c)
 \end{tabular}
\end{center}
\caption{Dynamo simulations with shallow meridional flow but with radial and latitudinal turbulent pumping included (same convention is followed as in Fig.~2). The toroidal (a) and poloidal field (b) plots at a phase corresponding to cycle maxima show the dipolar nature of the solutions, and the butterfly diagram at the base of the convection zone ($0.71 R_\odot$) clearly indicates the equatorward propagation of the toroidal field that forms sunspots.}
\end{figure}

\begin{figure}[!htb]
\begin{center}
\begin{tabular}{cc}
\includegraphics[scale=0.75]{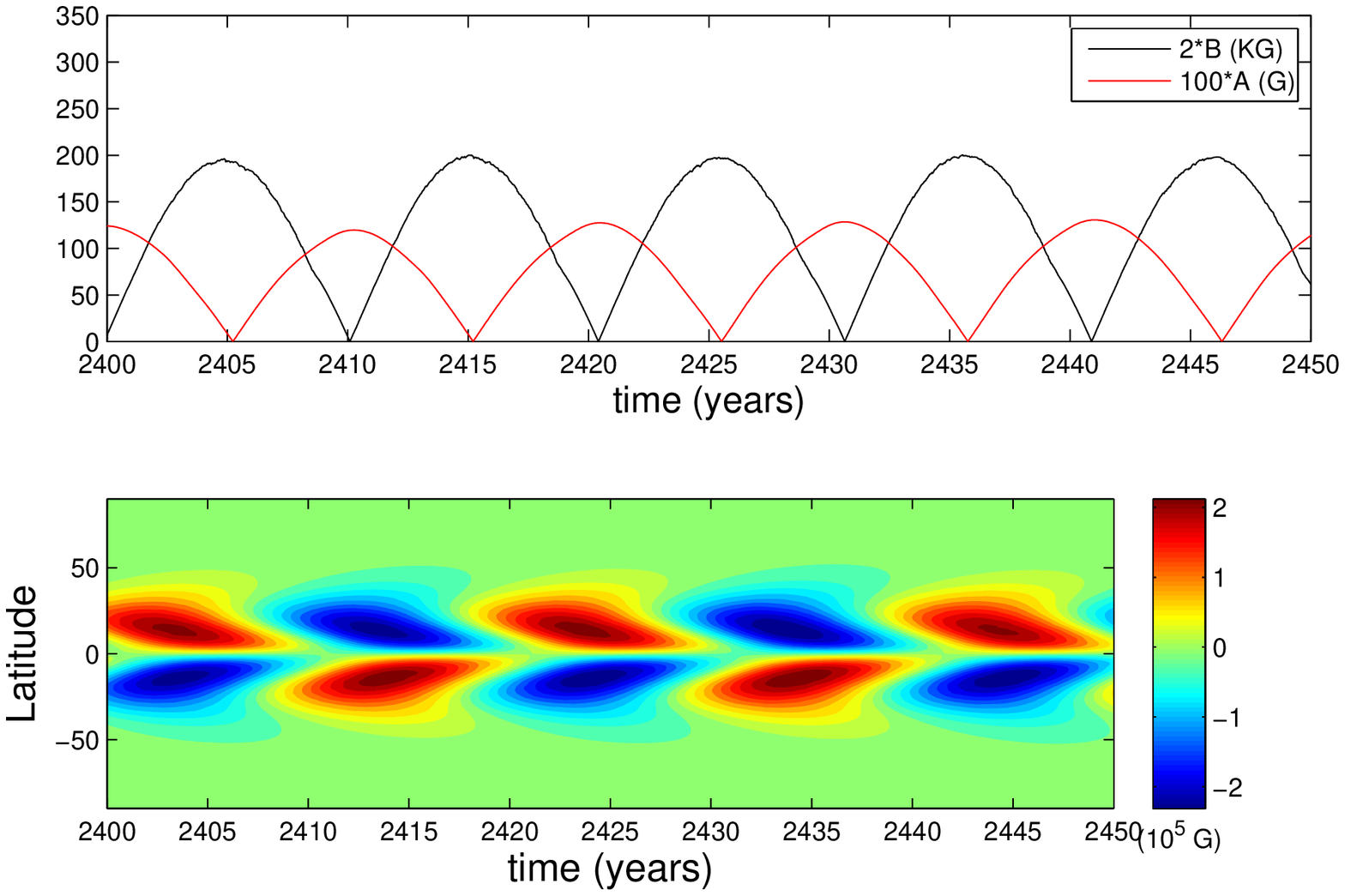}
 \end{tabular}
\end{center}
\caption{Dynamo simulations considering both shallow meridional flow and turbulent pumping but initialized with symmetric initial condition (quadrupolar state). Top panel shows the phase relationship between toroidal and poloidal field while bottom panel shows the butterfly diagram taken at the base of the convection zone ($0.71 R_\odot$).}
 \end{figure}
 
We now introduce both radial and latitudinal turbulent pumping in our dynamo model to explore whether a Babcock Leighton flux transport dynamo can operate with meridional flow which is much shallower than previously assumed; we also extend this study to the scenario where meridional flow is altogether absent.

The turbulent pumping profile is determined from independent MHD simulations of solar magnetoconvection \cite{ossen02, kapla06}. Profiles for radial and latitudinal turbulent pumping ($\gamma_r$ and $\gamma_\theta$) are:
\begin{eqnarray}
\gamma_r = -  \gamma_{0r} \left[ 1 + \rm{erf}\left( \frac{r - 0.715R_\odot}{0.015R_\odot}\right) \right] \left[ 1 - \rm{erf} \left( \frac{r-0.97R_\odot}{0.1R_\odot}\right) \right] \nonumber \\
\times \left[ \rm{exp}\left( \frac{r-0.715R_\odot}{0.25R_\odot}\right) ^2 \rm{cos}\theta +1\right] ~~~~
\end{eqnarray}
\begin{eqnarray}
\gamma_\theta = \gamma_{0\theta} \left[1+\mathrm{erf}\left(\frac{r-0.8R_\odot}{0.55R_\odot}\right)\right]
\left[1-\mathrm{erf}\left(\frac{r-0.98R_\odot}{0.025R_\odot}\right)\right]
\times \cos \theta \sin^4 \theta ~~~~
\end{eqnarray}
The value of $\gamma_{0r}$ and $\gamma_{0\theta}$ determines the amplitude of $\gamma_r$ and $\gamma_\theta$ respectively. Fig.~3 (top and bottom plot) shows that radial pumping speed (dashed lines) is negative throughout the convection zone corresponding to downward advective transport and vanishes below $0.7R_\odot$. The radial pumping speed is maximum near the poles and decreases towards the equator. Fig.~3 (top and bottom plot) shows that the latitudinal pumping speed (solid lines) is positive (negative) in the convection zone in the northern (southern) hemisphere and vanishes below the overshoot layer. This corresponds to equatorward latitudinal pumping throughout the convection zone.

Dynamo simulations with turbulent pumping generate solar-like magnetic cycles (Fig.~4 and Fig.~5). Now the toroidal field belt migrates equatorward, the solution exhibits solar-like parity and the correct phase relationship between the toroidal and poloidal components of the magnetic field (see Fig.~5). Evidently, the coupling between the poloidal source at the near-surface layers with the deeper layers of the convection zone where the toroidal field is stored and amplified, the equatorward migration of the sunspot-forming toroidal field belt and correct solar-like parity is due to the important role played by turbulent pumping. We note if the speed of the latitudinal pumping in on order of 1.0 ms$^{-1}$ the solutions are always of dipolar parity irrespective of whether one initializes the model with dipolar or quadrupolar parity. Interestingly, the latitudinal migration rate of the sunspot belt as observed is of the same order.
 \begin{figure}[!htb]
\begin{center}
\begin{tabular}{cc}
\includegraphics[scale=0.7]{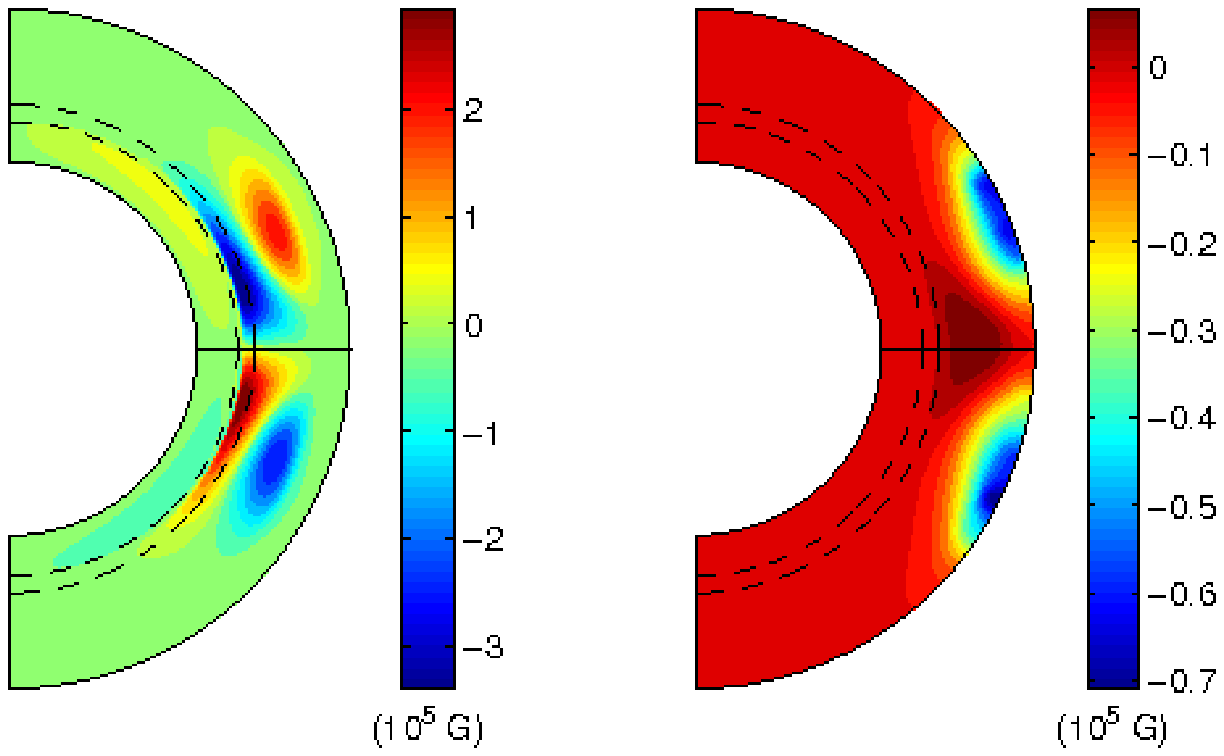} \\
(a)~~~~~~~~~~~~~~~~~~~~~~~~~~~~~~~~~~~~~~~~~~~~~~(b)\\
\includegraphics[scale=0.7]{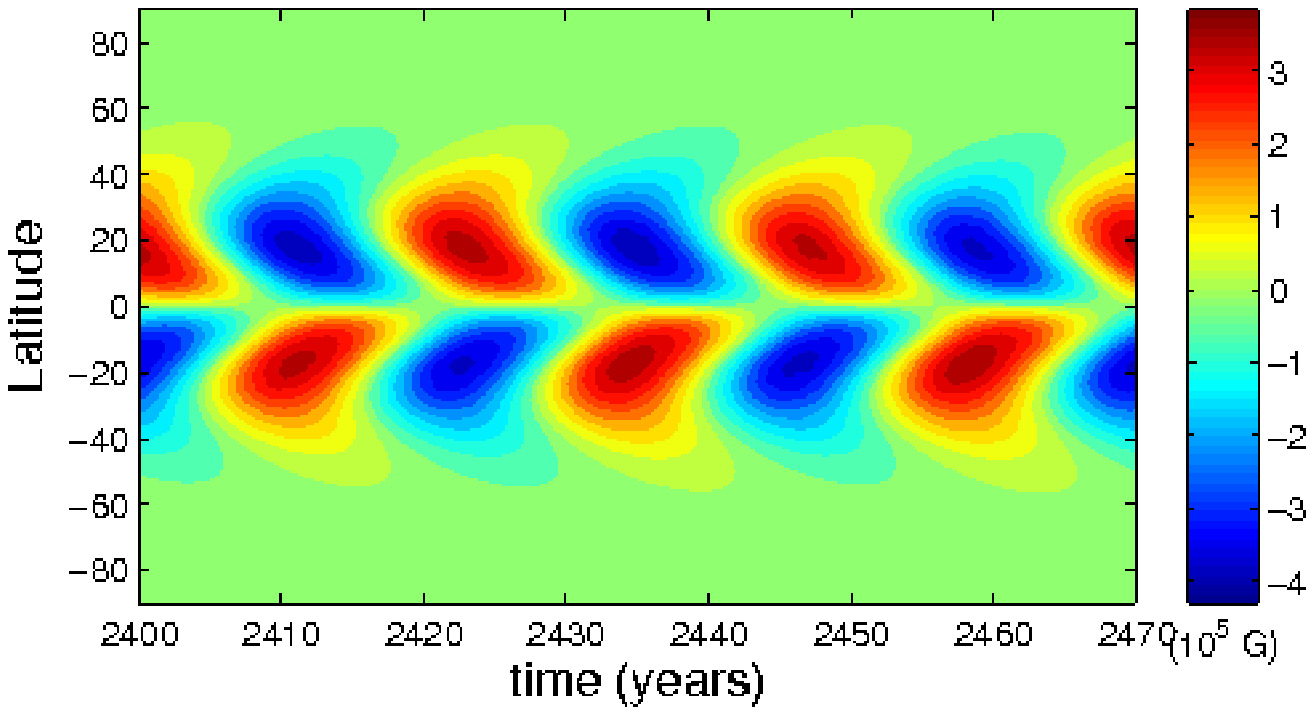} \\
~~~~(c)
 \end{tabular}
\end{center}
\caption{Results of solar dynamo simulations with turbulent pumping and without any meridional circulation. The convention is the same as in Fig.~2. The simulations show that solar-like sunspot cycles can be generated even without any meridional plasma flow in the solar interior. }
\end{figure}
The above result begs the question whether flux transport solar dynamo models based on the Babcock-Leighton mechanism that include turbulent pumping can operate without any meridional plasma flow. To test this, we remove meridional circulation completely from our model and perform simulations with turbulent pumping included. We find that this model generates solar-like sunspot cycles with periodic reversals (see Fig.~6) which are qualitatively similar to the earlier solution with both pumping and shallow meridional flow. However, we find that the surface magnetic field dynamics related to polar field reversal is limited to within 60 degrees latitudes in both the hemispheres. At higher latitudes (near the poles) the field is very weak and almost non-varying over solar cycle timescales. This is expected if the surface magnetic field dynamics is governed primarily by diffusion. Based on this result, we argue that this scenario of non-existent meridional circulation is not supported by current observations of surface dynamics which seem to suggest that the fields do migrate all the way to the poles.
\begin{figure}[!htb]
\begin{center}
\begin{tabular}{cc}
\includegraphics[scale=0.7]{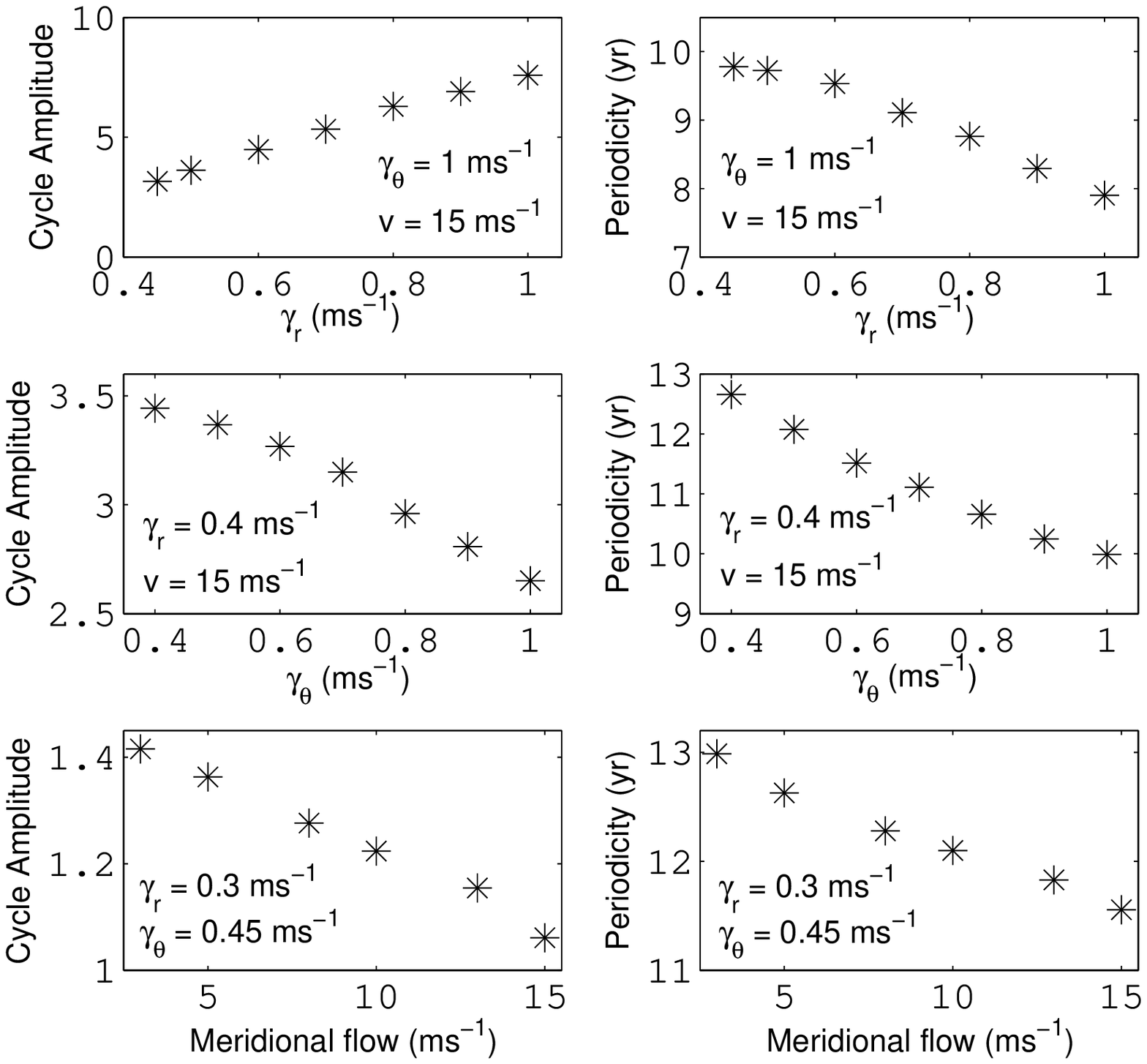} \\
 \end{tabular}
\end{center}
\caption{Dependence of amplitude and periodicity of simulated solar cycles on turbulent pumping (radial and latitudinal) and (shallow) meridional flow speeds. Pearson and spearman correlation coefficients are 0.99 and 1 respectively for top left plot and -0.99 and -1 respectively for all other plots.}
\end{figure}

Two important characteristics associated with the solar magnetic cycle are its amplitude and periodicity. While the periodicity of the cycle predominantly depends on the recycling time between toroidal and poloidal field, its amplitude depends on a variety of factors including dynamo source strengths and relative efficacy of transport timescales with respect to the turbulent diffusion timescale. We explore the dependency of the solar cycle period and amplitude to variations in the transport coefficients to explore the subtleties of the interplay between diverse flux transport processes. Figure 7 shows the dependency of cycle amplitude and periodicity on different velocity components like turbulent pumping and (shallow) meridional flow. A parametric analysis of this dependency yields the following relationships for cycle period ($T$) and cycle amplitude (Amp):

\begin{equation}
 T \simeq 9.7  ~ \gamma_r^{-0.25}  \gamma_\theta^{-0.26} v^{-0.068},
\end{equation}

\begin{equation}
 Amp \simeq 11.76 ~ \gamma_r^{1.07}  \gamma_\theta^{-0.27} v^{-0.16},
\end{equation}

which is gleaned from simulations within the following ranges: $0.25 ~ms^{-1} \leq \gamma_r \leq 1.25 ~ms^{-1}$, $0.25 ~ms^{-1} \leq \gamma_\theta \leq 1.25 ~ms^{-1}$ and $2 ~ms^{-1} \leq v \leq 15 ~ms^{-1}$; $\gamma_r$ and $\gamma_\theta$ are radial and latitudinal turbulent pumping speeds, and $v$ is the shallow meridional flow speed.

This analysis shows that cycle period and amplitude are both governed by diverse transport coefficients such as meridional flow speed, and radial and latitudinal components of turbulent pumping. As radial turbulent pumping carries the flux directly to the base of the convection zone where toroidal field is amplified, increase in the radial turbulent pumping speed leads to a decrease in cycle period. Increasing latitudinal pumping also has a similar effect on period which is similar to what is achieved by increasing meridional flow speed, namely a faster transport through the shear layer and thus shorter cycle periods. The cycle amplitude decreases on increasing the latitudinal pumping or meridional flow speed and this is due to the fact that less time is available for toroidal field induction when it is swept at a faster rate through the rotational shear layers. In surface flux transport models, a similar effect is found but due to a different reason -- wherein a faster meridional flow reduces the polar field strength because it takes flux of both polarity and deposits this at the poles (in effect carrying less net flux to the poles); in these simulations with a shallow meridional flow and the double-ring algorithm a similar mechanism could also be contributing to an overall reduction of the field strength. What is interesting to note though is the positive dependence of cycle amplitude on the radial pumping speed. We believe that a faster radial pumping moves the poloidal field down to the generating layers of the toroidal field in the deeper parts of the convection zone faster, thus allowing for less turbulent decay in the poloidal field strength; this eventually results in a stronger poloidal field in the SCZ which generates a stronger toroidal component.

We note that the derived exponents for the cycle period above differ from that determined by Guerrero \& de Gouveia Dal Pino (2008). The cycle period in our simulations is more strongly dependent on the latitudinal speed of turbulent pumping and less so on meridional circulation, whereas in Guerrero \& de Gouveia Dal Pino (2008) it is the exact reverse. In our model the meridional flow is very shallow and limited to only the top $10 \%$ of the SCZ, whereas in the model setup of Guerrero \& de Gouveia Dal Pino (2008), the meridional flow penetrates down to about 0.8 $R_\odot$; this we believe makes their dynamo cycle periods more sensitive to meridional flow as compared to latitudinal pumping.

Generally, we find solar-like solutions in a modest turbulent pumping speed range on the order of 1 ms$^{-1}$. This parameter study shows that our result are robust to reasonable variations in turbulent pumping coefficients and also points to how the latter may determine solar cycle strength and periodicity.

\begin{figure}[!htb]
\begin{center}
\begin{tabular}{cc}
\includegraphics[scale=0.7]{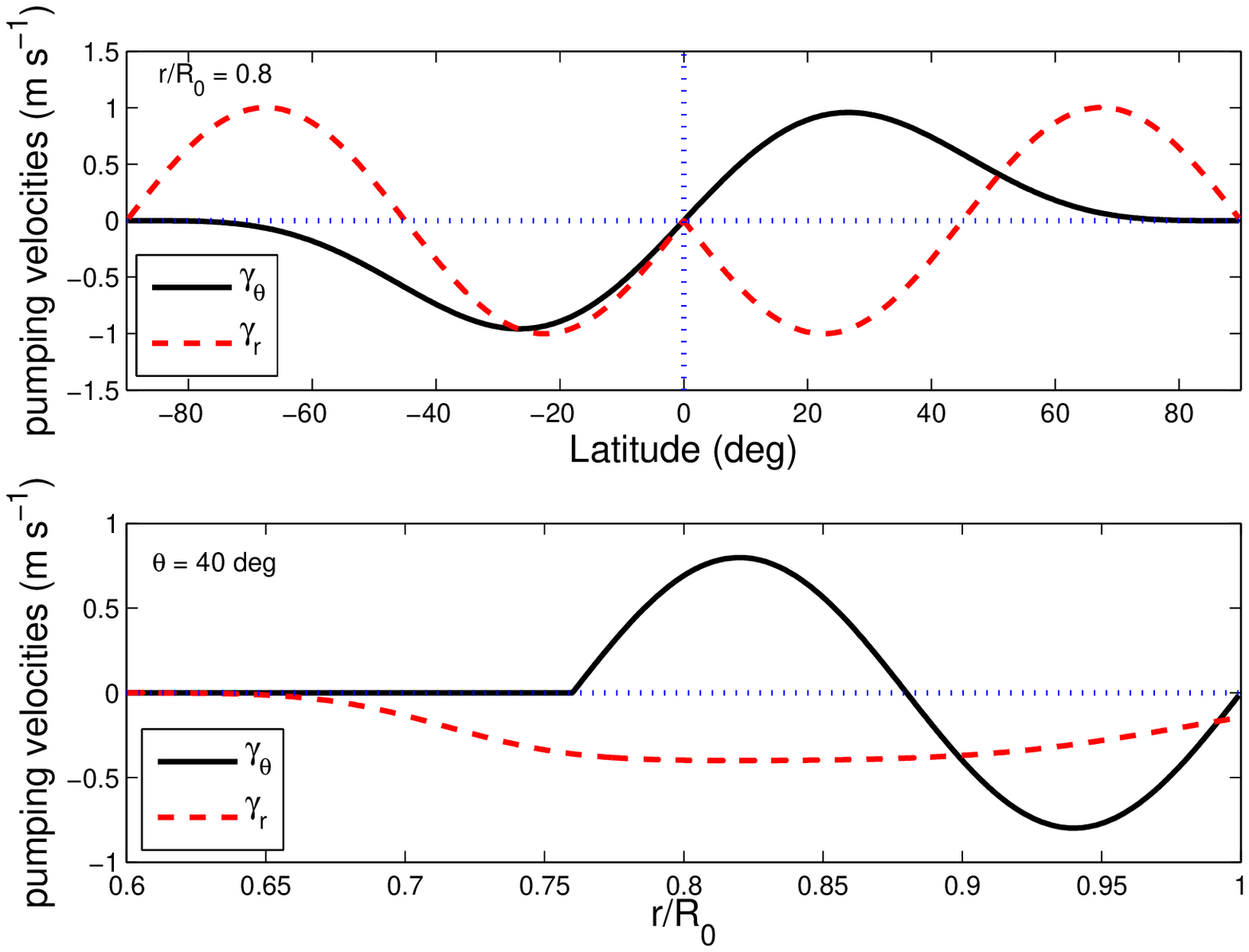} \\
\includegraphics[scale=0.8]{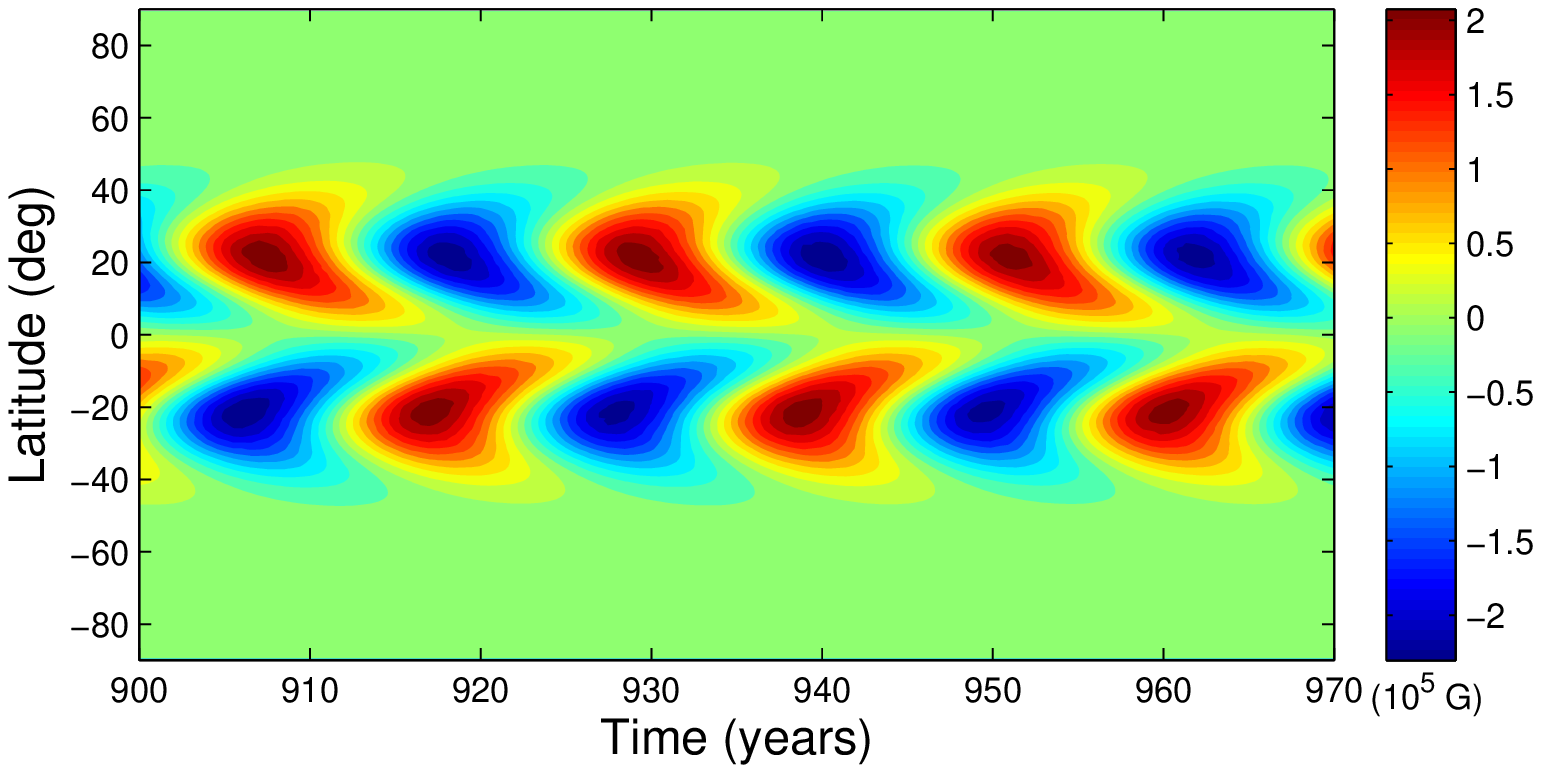} \\
 \end{tabular}
\end{center}
\caption{Results of solar dynamo simulations (with shallow meridional flow) utilizing an alternate and more complex turbulent pumping profile based on Warnecke et al. (2016). First two plots show the radial and latitudinal variation of turbulent pumping generated by analytic approximations to the Warnecke et al. (2016) results. The butterfly diagram (bottom plot) taken at the base of the convection zone ($0.71 R_\odot$) in our dynamo simulations indicate that solar-like solutions are reproduced with this alternative profile.}
\end{figure}
As there is some uncertainty regarding the exact details of turbulent pumping profiles, we have tested an alternative turbulent pumping profile based on Warnecke \textit{et al.} (2016). Recent magnetoconvection simulations performed by Warnecke et al. (2016) suggest that radial pumping is downward throughout the convection zone below $45^{\circ}$ and upward above $45^{\circ}$, while latitudinal pumping is poleward at the surface and equatorward at the base of the convection zone. Our generated turbulent pumping profiles in the northern hemisphere (defined within $0 \leq \theta \leq \pi/2$) based on the suggestions of Warnecke \textit{et al.} (2006) are:
\begin{eqnarray}
\gamma_r = -  \gamma_{0r} \left[ 1 + \rm{erf}\left( \frac{r - 0.715R_\odot}{0.05R_\odot}\right) \right] \left[ 1 - \rm{erf} \left( \frac{r-0.98R_\odot}{0.08R_\odot}\right) \right] \nonumber \\
\times \sin(4 \theta), ~~~~
\end{eqnarray}
\begin{eqnarray}
\gamma_\theta = \left\{\begin{array}{cc}
\gamma_{0\theta} \sin \left[\frac{2 \pi (r- R_p)}{R_0-R_p}\right]
\times \cos \theta \sin^4 \theta ~~~~  & r \geq R_p\\
  0  & r<R_p
\end{array}\right.,
\end{eqnarray}

where $R_p= 0.76R_\odot$ i.e. the penetration depth of the latitudinal pumping. The amplitudes of $\gamma_r$ and $\gamma_\theta$ are determined by the value of $\gamma_{0r}$ and $\gamma_{0\theta}$ respectively. Turbulent pumping profiles in the southern hemisphere are generated by replacing colatitude $\theta$ by $(\pi- \theta)$. Fig.~8 (the top and middle panels) show that our generated turbulent pumping profiles capture the basic essence of the suggestions made by Warnecke \textit{et al.} (2016). Our simulations (with shallow meridional flow) and the more complex turbulent pumping profile gleaned from Warnecke \textit{et al.} (2016) reproduce broad features of the solar cycle and are qualitatively similar to those detailed earlier.

\section{Discussions}

In summary, we have demonstrated that flux transport dynamo models of the solar cycle based on the Babcock-Leighton mechanism for poloidal field generation does not require a deep equatorward meridional plasma flow to function effectively. In fact, our results indicate that when turbulent pumping of magnetic flux is taken in to consideration, dynamo models can generate solar-like magnetic cycles even without any meridional circulation although the surface magnetic field dynamics does not reach all the way to the polar regions in this case. Our conclusions are robust across a modest range of plausible parameter space for turbulent pumping coefficients and also indicate some tolerance for diverse pumping profiles.

These findings have significant implications for our understanding of the solar cycle. First of all, the serious challenges that were apparently posed by observations of a shallow (and perhaps complex, multi-cellular) meridional flow on the very premise of flux transport dynamo models stands resolved. Turbulent pumping essentially takes over the role of meridional circulation by transporting magnetic fields from the near-surface solar layers to the deep interior, ensuring that efficient recycling of toroidal and poloidal field components across the SCZ is not compromised. While these findings augur well for dynamo models of the solar cycle, they also imply that we need to revisit many aspects of our current understanding if indeed meridional circulation is not as effective as previously thought. For example, our simulations indicate that variations in turbulent pumping speeds can be an effective means for the modulation of solar cycle periodicity and amplitude.
 
It has been argued earlier that the interplay between competing flux transport processes determine the dynamical memory of the solar cycle governing solar cycle predictability \cite{yeat08}. If turbulent pumping is the dominant flux transport process as seems plausible based on the simulations presented herein, the cycle memory would be short and this is indeed supported by independent studies \cite{kar12} and solar cycle observations \cite{munoz13}. It is noteworthy that on the other hand, if meridional circulation were to be the dominant flux transport process, the solar cycle memory would be relatively longer and last over several cycles. This is not borne out by observations.

Previous results in the context of the maintenance of solar-like dipolar parity have relied on a strong turbulent diffusion to couple the Northern and Southern hemispheres of the Sun \cite{chat04}, or a dynamo $\alpha$-effect which is co-spatial with the deep equatorward counterflow in the meridional circulation assumed in most flux transport dynamo models \cite{dikp01}. However, our results indicate that turbulent pumping is equally capable of coupling the Northern and Southern solar hemispheres and aid in the maintenance of solar-like dipolar parity. This is in keeping with earlier, independent simulations based on a somewhat different dynamo model \cite{guer08}.

Most importantly, our results point out an alternative to circumventing the Parker-Yoshimura sign rule constraint \cite{park55, yosh75} in Babcock-Leighton type solar dynamos that would otherwise imply poleward propagating sunspot belts in conflict with observations. Brandenburg {\it et al.} (1992) and Ossendrijver {\it et al.} (2002) had already pointed towards this possibility in the context of mean-field dynamo models. While a deep meridional counterflow is currently thought to circumvent this constraint and force the toroidal field belt equatorward, our results convincingly demonstrate that the latitudinal component of turbulent pumping provides a viable alternative to overcoming the Parker-Yoshimura sign rule in Babcock-Leighton models of the solar cycle (even in the absence of meridional circulation).

We note however that our theoretical results should not be taken as support for the existence of a shallow meridional flow, rather we point out that flux transport dynamo models of the solar cycle are equally capable for working with a shallow or non-existent meridional flow, as long as the turbulent pumping of magnetic flux is accounted for; this is particularly viable when turbulent pumping has a dynamically important latitudinal component. Taken together, these insights suggest a plausible new paradigm for dynamo models of the solar cycle, wherein, turbulent pumping of magnetic flux effectively replaces the important roles that are currently thought to be mediated via a deep meridional circulation within the Sun's interior. Since the dynamical memory and thus predictability of the solar cycle depends on the dominant mode of magnetic flux transport in the Sun's interior, this would also imply that physics-based prediction models of long-term space weather need to adequately include the physics of turbulent pumping of magnetic fields.

\begin{acknowledgments}

We acknowledge the referee of this manuscript for useful suggestions. We thank J\"orn Warnecke for helpful discussions related to the adaptation of the turbulent pumping profile from Warnecke {\it et al.} 2016.  We are grateful to the Ministry of Human Resource Development, Council for Scientific and Industrial Research, University Grants Commission of the Government of India and a NASA Heliophysics Grand Challenge Grant for supporting this research.
\end{acknowledgments}

\end{document}